\documentclass[letterpaper,12pt,oneside,onecolumn]{article}%
\usepackage{geometry}%
\usepackage{titlesec}%
\titleformat{\section}[hang]{\normalfont\normalsize\bfseries}{\thesection}{12pt}{\centering}%
\titleformat{\subsection}[display]{\normalfont\normalsize}{\thesubsection}{12pt}{\underline}%
\titleformat{\subsubsection}[runin]{\normalfont\normalsize}{\thesubsubsection}{12pt}{\underline}%
\newcommand{\TPCI}[3]{%
\begin{flushright}%
    \begin{scriptsize}%
        \textbf{#1} \textit{#2} \\
        \textbf{\textit{TMS (The Minerals, Metals \& Materials Society), #3}}\\%
    \end{scriptsize}%
\end{flushright}%
}%
\newcommand{\PaperTitle}[1]{%
\begin{center}%
    \begin{large}%
        \textbf {#1} \\%
    \end{large}%
\end{center}%
}%
\newcommand{\AuthorList}[1]{%
\begin{center}%
    {#1} \\%
\end{center}%
}%
\newcommand{\AuthorAffiliations}[1]{%
\begin{center}%
    {#1} \\%
\end{center}%
}%
\newcommand{\Keywords}[1]{%
\begin{center}%
   Keywords: {#1} \\%
\end{center}%
}%

\usepackage{graphics}
\usepackage{epsfig}
\begin{document}%
\TPCI{Title of Publication}{Edited by}{Year}%
\PaperTitle{Atomistic modelling of the Shape Memory Effect}
%
%
%Author(s) Name(s), Author(s) Affiliation(s), and Keywords%
%
\AuthorList{G.J.Ackland}%
\AuthorAffiliations{School of Physics and Centre for Science at
Extreme
Conditions, The University of Edinburgh, Mayfield Road, Edinburgh, EH9 3JZ, UK}%
\section{Abstract}

This paper reviews the status of molecular dynamics as a method in
describing solid-solid phase transitions, and its relationship to
continuum approaches\cite{christ,Bhattacharya}. Simulation work done in NiTi and Zr
using first principles and semi-empirical potentials is presented.
This shows failures of extending equilibrium thermodynamics to the
nanoscale, and the crucial importance of system-specific details to
the dynamics of martensite formation.  The inconsistency between experimental 
and theoretical crystal structures in NiTi is described, together with its 
possible resolution in terms of nanoscale effects.

\section{Introduction}

This paper reviews work done on atomistic modelling of martensitic
phase transitions, and discusses the unresolved issues in linking
various methods of modelling solid-solid phase transitions.  In
addition to discussing phase transitions in NiTi and Zr, following the
spirit of the meeting, I attempt to review and highlight what is not
widely known or properly appreciated.  My benchmark for "what is
known" is primarily based on actual discussions with other scientists,
regardless of whether it has appeared in the literature.

The molecular dynamics method is in principle the most versatile way
of describing solid-solid phase transitions.  The crystal and
interfacial structures arise automatically from the interatomic
potential, as are the long range strain effects and there is no need
for the implicit assumptions about microscopic detail and symmetry
entailed in continuum methods.  In particular, thermodynamics emerges
from molecular dynamics rather than being an input, so all the
fluctuations are incorporated properly.

The downside of molecular dynamics is twofold - the models for
interatomic forces are unreliable and the timestep is very short
because it is set by the phonon vibrational frequencies which
typically not relevant to nucleation and diffusive processes.

It may appear that methods which access longer timescales and larger
systems, such as kinetic monte carlo or phase fields are more useful,
however this appearance is misleading.  These methods have conspicuous
success in certain systems where their assumptions are valid, but
cannot be applied in general.

Thus, molecular dynamics is an essential part of any attempt to model
real materials.  In this paper we discuss a number of issues and
questions which have arisen from our work in molecular dynamics, but
which tend to be ignored or glossed over in other techniques.  Along
the way, I discuss a few points which appear to have been ignored in
much of the literature, highlight some of the current unresolved
issues and include the consensus of many discussions at the meeting in
Phoenix.

Throughout, we discuss only structural transitions, assuming the
chemical composition to be constant.  The transitions may then be
driven either by stress or temperature.

Complete atomistic modelling of the shape memory effect has not yet
been done.  It has been delayed primarily by the lack of appropriate
potentials for the interesting materials such as NiTi.  It is likely
that some information about the mechanisms could be obtained for
Parrinello-Rahman dynamics with very simple potentials which describe
the B2-B19 phase transition on cooling, however efforts to date has
concentrated on the B2-B19' transition exhibited by real shape memory
alloys, such as NiTi, and hcp-bcc in elements such as Zr and Ti, which 
do not exhibit the shape memory alloy..

The theory of the shape memory effect is straightforward: the material
undergoes a martensitic transition on cooling to a lower symmetry
structure, in which there is a one to one correspondence between atoms
in each phase.  The set of lower symmetry structures which preserve
this unique relationship to the austenite is referred to as the
Eriksen-Pitteri neighbourhood (EPN) of the austenite.  Deformation of
the martensite occurs by twin boundary motion, again preserving the
EPN.  The standard theory for shape memory alloys required that at no
point in the transformation or deformation of the structure do the
atoms leave this EPN.  A minimum constraint here is that the transition is displacive, not diffusive\cite{DD}

There is also a two-way shape memory effect, in which the
high-symmetry austenite phase stores a memory of the low-symmetry
martensite phase.  This effect is not usually so pronounced, and
cannot be explained in terms of EPN - a high symmetry phase cannot lie
in the EPN of the low symmetry one.

\section{Applicability of Thermodynamics and Statistical Mechanics}

Avogadro's is one of the larger numbers, but it isn't always infinite

Thermodynamics is a continuum limit approximation to statistical
mechanics.  Statistical mechanics is a probabilistic approximation to
the chaotic dynamics of particles.  At base, everything is made of
atoms and electrons.

Gibbs' free energy is thus an emergent property of particle dynamics.
In bulk materials, it is an exceptionally good one - when averaging
over say 1022 atoms the statistical errors are around 10-11, so for
any practical purpose we can treat thermodynamics as an exact theory
for bulk phases at equilibrium.  The notion of a driving force based
on a differential of a free energy is less well motivated since the
existence of a gradient means that the system is out of equilibrium.
Strictly speaking, the Maxwell-Boltzmann distribution underlying
statistical mechanics then does not apply, and so nor does equilibrium
thermodynamics.  Again, we are saved by a division of time and
lengthscales.  The entropic contribution in crystalline solids is
contained in phonon vibrations, whose frequency is many orders of
magnitude greater than the typical timescales of a phase transition.
Similarly, the typical lengthscales involved in microstructures are
several orders of magnitude larger than interatomic spacings.  Thus we
can usually regard local regions as being at equilibrium because a
sufficiently large number of atoms are vibrating for a sufficiently
long time for the approximations underlying thermodynamics to be
valid.

The movement of interfaces driven by a free energy difference is
similarly ill-defined.  One can envisage two limits, in one the
interface moves freely back and forth on a timescale much faster than
its net drift.  This is a promising case for using free energy
differences, since the system is sampling both phases efficiently.  In
other cases the interface is sessile - unable to move thermally, and
tends to advance by a stick-slip process.  In neither case, however,
is the rate of motion of the interface given by the free energy
gradient - one has to introduce some arbitrary timescale.  In simple
cases, where only a few types of interfaces are present, there may be
only one timescale, so the dynamics of the transition can be captured
correctly.  In more complex cases where some boundaries are sessile
and others mobile, the transition kinetics and resultant
microstructure will be primarily determined by boundary mobility
rather than bulk free energies.

In continuum modelling of martensites, the interface sets an energy,
and hence length, scale.  In real materials the interfacial energy
varies strongly with angle, and has cusps at special values (so called
low sigma boundaries).  Normally they are atomically sharp, which
presents a further challenge

Thus methods based directly on thermodynamics will become a problem in
very fast transitions (e.g. shock waves on martensites) or extremely
fine microstructures (the canonical example being a glass
transition\cite{glass}.

\section{Invariant Plane Interfaces}

The interfacial plane between two solid phases of the same
composition, with one having transformed from the other, may be
uniquely determined if the transition mechanism is known.  Where the
transition mechanism is assumed to be purely displacive, the
interfacial plane is purely determined by crystalline geometry - the
invariant plane. This is defined by taking a slice through the unit
cell of each crystal such that the two faces are identical in area and
shape.  

The requirement for the transformation to occur on an invariant plane
has some microstructural consequences:
any volume change must be absorbed by contraction of the cell 
normal to the interface 
plane, this generates a strain energy around a nucleating region: 
hence growth in this direction is hindered and martensite typically grows as 
a plate.  In general, the invariant plane is at an arbitrary angle to the 
crystallographic directions.

The invariant plane should not be confused with the transformation
mechanism.  This determines how the atoms get from their positions in
one crystal structure to their positions in the other.  After the
material is fully transformed, it is the transformation mechanism 
which determines the orientation of one grain to another.

\subsection{Viewing the microstructure}

It is essential for analysis of the MD that some we have some way of
assigning individual atoms to particular twins or crystal structures.
This process is rather arbitrary, but a good method is to use the
distribution of cosines of angles to nearest neighbours to distinguish 
structures.  This allows one to determine fcc, hcp and bcc regions of a 
crystal, and at the same time determine the orientation of the structures.\cite{may1}

\section{Atomistic simulation of martensite - zirconium}

Atomistic simulation gives the possibility of studying martensitic
phase transitions without any of the assumptions required in the
continuum approaches.  The interatomic potential used determines
whether the simulation reflects any particular material, however the
topological and geometric features arising will be generic.  From a
number of simulations carried out on zirconium, the detailed structure
of the twin boundary emerges as the crucial feature for the dynamics.
Details of boundary structure is generally ignored in the continuum or
phase field approaches, at best boundaries are assumed to have some
constant energy or behaviour irrespective of orientation.  However,
for shape memory effect the boundary must be capable of moving without
leaving the EPN the austenite.  
This places constraints on the mechanism by which
the boundary can move.

\section{Pretransition phenomena and intermediate phases}

\begin{figure}[ht]
\protect{\includegraphics[width=0.55\columnwidth]{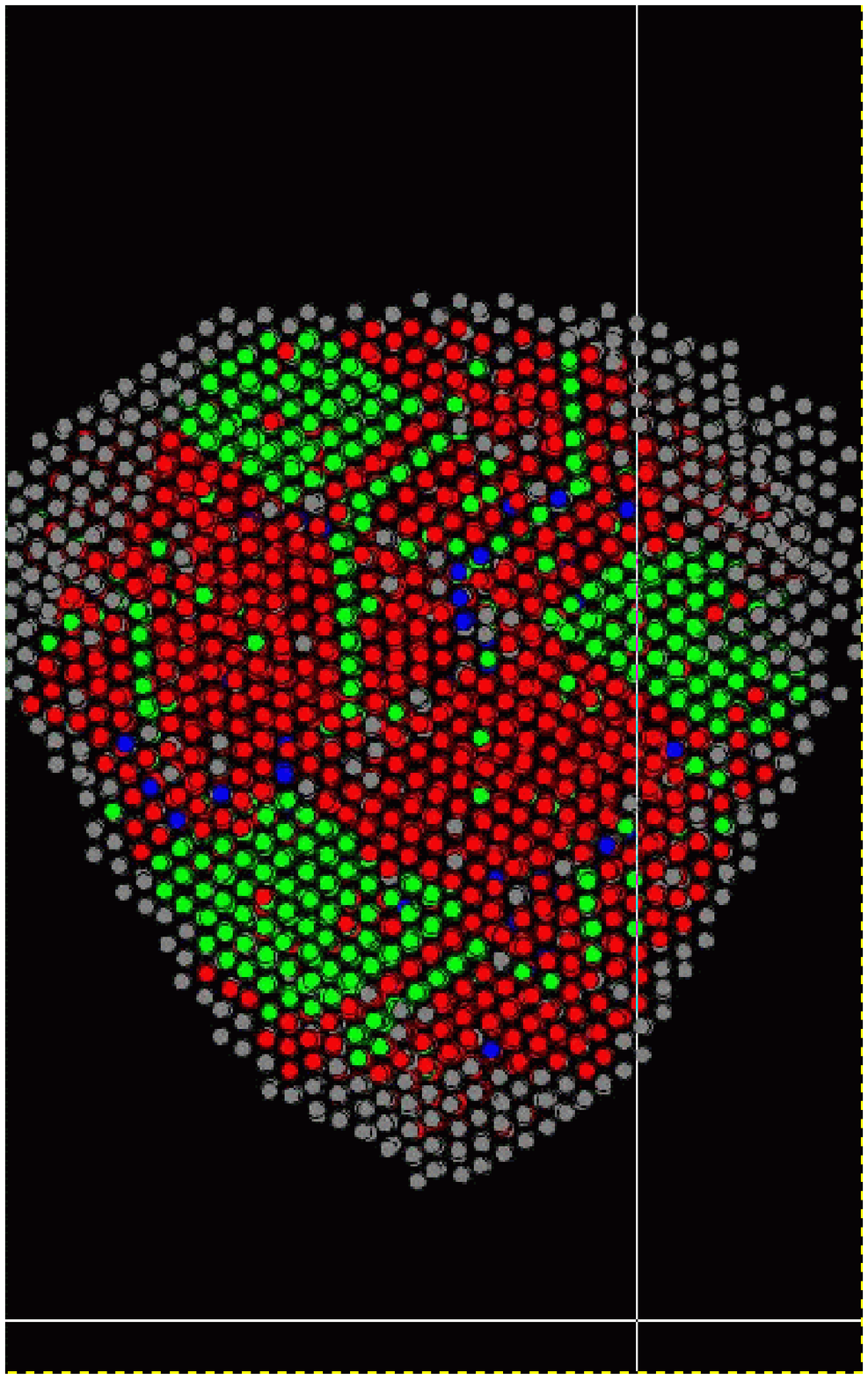}}
\protect{\includegraphics[width=0.45\columnwidth]{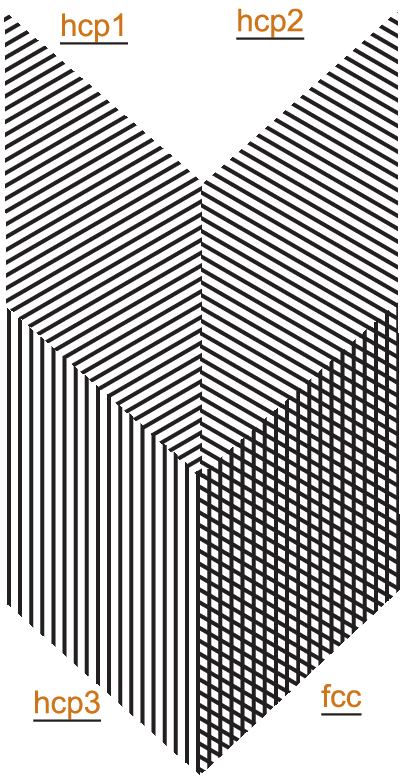}}
\caption{(left) Snapshot of a slice through supercooled, dynamically
stabilised bcc zirconium, showing instantaneous atom positions and
allocating local crystal structure (red=hcp, green=fcc, blue=bcc,
grey=boundary) according to method of Jones and Ackland\cite{Jones}.
Single lines of green represent stacking faults\cite{nab} in the hcp,
e.g.ABABCBCB stacking.  (right) Schematic representation of this type
of structure.  Lines show close packed planes viewed end on, which
exist in more than one plane in fcc, allowing the stacking to continue
unbroken across the interface.  None of the arrangements hcp1-3 can
replace fcc without introducing a high energy boundary or a variant which spans the simulation (and is therefore subject to large stress \protect\label{Joneszr}}
\end{figure}

In transitions from bcc via the Nishiyama-Wassermann mechanism the
(110)$_{bcc}$ planes convert to basal (0001)$_{hcp}$ planes in hcp.
Different hcp variants arise from different original sets of
(110)$_{bcc}$.  This means that the variants must be rotated with
respect to each other by 30 or 60 degrees.  If one investigates the
high temperature crystal structure of Zr on cooling from the bcc phase
prior to the transition, the structure is found to fluctuate\cite{may2}, a
snapshot being shown in fig\ref{Joneszr}.

In this figure we see the effects of stress-induced preferential
nucleation.  In the bcc-hcp transformation, the strain along the
(111)$_{bcc}$ direction (a line of near neighbours) is small
(perpendicular to the plane) thus once the materials begins to
transform to an hcp variant, and large stress filed is developed in
the plane, which induces a transformation into the other two variants
which share the particular (111)$_{bcc}$ plane.  Thus the stress 
nucleates cp basal planes from the  
$(01\overline1)_{bcc}, (1\overline10)_{bcc}$ and $(101\overline)_{bcc}$
planes.  This rapid nucleation gives rise to a columnar microstructure  fig\ref{Joneszr}.

However, these variants cannot tile a plane with all interfaces being
60 degrees.  In this nanostructured material, the large boundary
energy of the 30 degree twin is sufficient to prevent a microstructure
comprising only these three variants.  What appears is a region of
fcc, stabilised by the low energy of the interface between it and the
adjacent hcp (the interface is simply a basal stacking).

The existence of metastable crystal phases close to the martensitic
transition is reported in many systems, notably the R phase in NiTi.
They are usually referred to as ``intermediate'' phases, often with
the implication that they are thermodynamically stable over a small
region.  This simulations suggests another picture, the
``intermediate`` phases (R in NiTi, fcc here) actually are stabilised
at the nanoscale both by the strain field of the precritical hcp phase
nuclei and their very low interfacial energy. As the structure
coarsens, they become metastable but their elimination may be
kinetically hindered.  The existence of such metastable phases could
play a important role in compensating the strain field, and delaying
the full transformation to martensite.

\begin{table}[t]
\begin{center}

\begin{tabular}{|c|cc|cc|}
\hline
  k &$ A_k$ $(eV/\AA^3 )$     & $R_k$ $ (\AA )$ &   $a_k$  $(eV/\AA^3 )$& $r_k
$  $ (\AA )$ \\
\hline 
 1 & -0.61248219 &  5.5763004 & 0.50569395   & 5.5763004  \\
 2 &  0.87645549 &  5.4848856 & -0.00890725 & 4.7992749  \\
 3 & -0.21947820 &  5.2106413 & -  & -  \\
 4 & -0.01371379 &  4.3422011  & - & -  \\
 5 &  0.68830444 &  3.6565904   & - & -  \\
 6 &  1.45995293 &  3.1995166   & - & -  \\
\hline
&
$ E_i = \frac{1}{2}\sum_{j}V(r_{ij})\ - \sqrt{\rho_i} $\\

&$ V(r) = \sum_{k=1}^2 a_k(r_k - r)^3 H(r_k - r)$\\
&
$\rho_i = \sum_{j}\phi (r_{ij}) = \sum_j\sum_{k=1}^6 A_k(R_k - r)^3 H(R_k - r)
$\\
\hline\end{tabular}
 \label{tab:coeff}
\end{center}
\caption{Energy function and parameters for zirconium potential where  $H(x)$ is the Heaviside step function, $r_{ij}$ is the separation
between atoms $i$ and $j$ 
}
\end{table}

\section{Twins and intervariant boundaries}

The angle between martensitic twins formed by cooling from an
austenite is fully determined by crystal structure and transition
mechanism.  There is no reason for it to be a low energy boundary.  In
the case of zirconium it turns out that the 30 degree boundary is very
high energy and is not seen in the simulations.  This restricts the 
variants which can be observed.

It is well established that the strain minimising microstructure is a
laminate of twins, and the boundary energy difference means they are
always 60 degree tilt this is clearly seen in the simulations.  A
$60\circ$ tilt is close to the a $61.5^\circ$ symmetric tilt boundary
of the $(10\overline{1}1)$ twin, a cusp in the energy vs angle graph
and with low boundary energy (0.58mJm$^{-2}$ according to this
potential). For an isolated twin boundary, the discrepancy in angle is
made up by twinning dislocations, however, when the twins are small
these dissociate into partial twinning dislocations with a basal
stacking fault extending through the twins\cite{wrongstack}.

\begin{figure}[ht]
\protect{\includegraphics[width=\columnwidth]{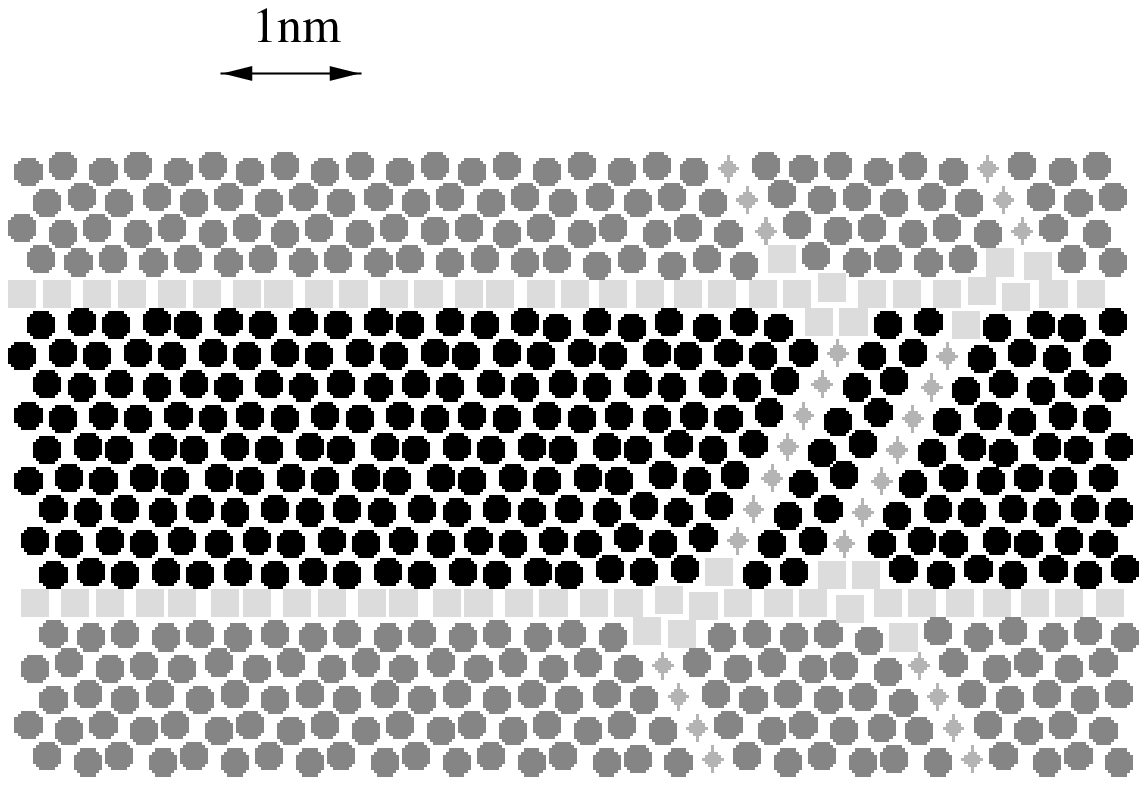}}
\caption{ Structure of the 60 degree twin boundary in zirconium.  The
position of the twin boundary partial dislocations are made clear by the
stacking fault emanating from it (circles: hcp, squares: boundary:
crosses stacking fault) Figure created by U.Pinsook \protect\label{maytwin}}
\end{figure}

The structure is shown in Figure\ref{maytwin}.  The density of 
twin boundary partial dislocations is determined by crystal 
geometry (in particular, the c/a ratio of hcp zirconium).  Thus to simulate a real material, using a potential which reproduces the lattice parameters accurately is crucial.

\section{Twin boundary motion}
As explained elsewhere in these proceedings\cite{pond} a twin boundary
dislocation is free to glide in the twin only if its Burgers vector
lies in the twin plane.  This is not the case for the twin boundary
partial dislocations in Figure\ref{maytwin}.  Thus the twin boundary 
cannot move in a diffusionless manner by motion of these twin boundary
partial dislocations.

The boundary does move, however, in response to external
stress\cite{may3}.  The role of twin boundary partial dislocations is
to act as stress concentrators which cause nucleation of other twin
boundary dislocation which are mobile in the boundary.  The mechanism
is similar to that observed for the interaction of a crystal
dislocation with a twin boundary \cite{serra}.  In the absence of the 
twin boundary partial dislocations, there is no source for mobile dislocations and so a perfect  $(10\overline{1}1)$ twin cannot move\cite{may3}

\section{NiTi}

NiTi is the canonical example of a shape memory alloy.  Ideally one
would like to do calculations with a potential for NiTi, which
described the high-T B2 phase, the low-T B19' phase and the metastable 
R phase.  In this section we discuss the problems encountered in 
generating such a potential.

\begin{figure}[ht]
\protect{\includegraphics[width=0.8\columnwidth]{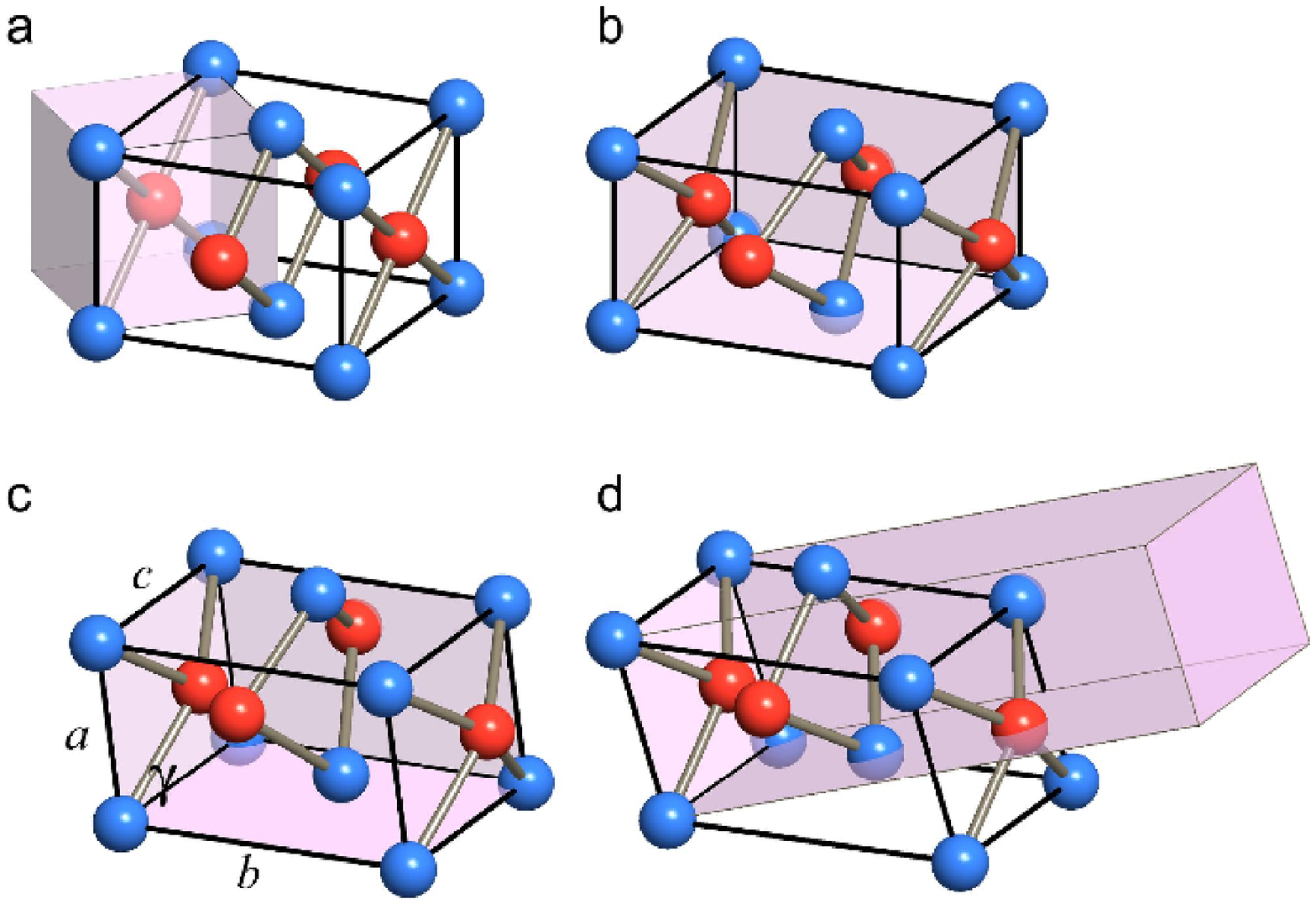}}
\caption{ Crystal structures in NiTi, and their relation via the
soft-phonon mode transformation mechanism (a) doubled B2 structure
oriented along (110) - here and elsewhere the conventional unit cell
is shown by shading.  (b) B19 structure obtained by tetragonal
deformation of B2, plus shuffle of internal coordinates (c) monoclinic
distortion of B19 to experimentally reported NiTi B19' martensite
structure. (d) increased monoclinic angle to DFT minimum energy
structure, B33, with additional tetragonal symmetry as shown.
\cite{xhuang}\protect\label{NiTiXtal}}
\end{figure}

The B2-B19' phase transition observed in NiTi can be simply understood
in terms of the Nishiyama-Wassermann-type mechanism observed in
zirconium (see fig.\ref{NiTiXtal}.  If one ignores the difference
between Ni and Ti atoms, B2 is simply bcc, the different species do
not disrupt the cubic symmetry.  B19 has the same symmetry as hcp,
except that the two species break hexagonal symmetry making the
structure to tetragonal.  B19' is a small distortion from B19.  We
might thus hope that similar potentials to those used in zirconium
would work for NiTi.

\subsection{Interatomic potentials for NiTi}

Experimental data for stress-free single crystal NiTi martensite is
very hard to obtain. Crystallography has been done on samples prepared
by strain-induced transformation above the martensite start
temperature, followed by quenching to ambient condition\cite{kudoh}.  It is
unclear whether such crystals are strain-free. In the absence of
detailed calorimetry for the phase transition energies, the standard
way to proceed in making potentials is to generate a database of ab
initio energies.  There are many free parameters in the B19'
structure, and great care has to be taken to ensure the true minimum
is found\cite{xhuang}.  When this is done, it turns out that the DFT
minimum energy structure is different from that found in the
experiment - the $\gamma$ angle in B19' is 107\cite{xhuang} rather than 98 degrees\cite{kudoh}.
Moreover, 107 degrees is exactly the right angle to create a
doubled unit cell with tetragonal symmetry - the B33 structure.
 
This conflict between DFT and experimental energetics makes it
difficult to attempt to fit a potential to NiTi combining data from 
each.  This explains why little progress has been made on the problem to date.

\subsection{Ab initio calculations in NiTi}

One way to avoid the problem with lack of accurate potentials is to
use ab initio molecular dynamics DFT calculations for direct
calculations of the microstructure.  We have carried out density
functional calculations based on pseudopotentials and plane waves
using the VASP code, of up to 100 atoms in the B2 phase, cooling using
a Nose thermostat from 400-200K, then quenching to 0K for analysis.
Two supercell geometries were tried, in one the cell was orientated
along (100), (0$\overline{1}$1), (011) requiring 64 atoms, while in the other
(0$\overline{1}$1), ($\overline{1}$11) (211) with 96 atoms was used.  The
classical work with zirconium indicated that this may prove too small,
and certainly finite size effect will be significant, and the 96 atom cell attempted to ease that problem being only 5A thick in the z-direction

On cooling the structures dropped out of the bcc phases.  We used the
angle based analysis\cite{Jones} to assign hcp, fcc or bcc crystal
structures. For a displacive transition, these would pick up local
B19, L1$_0$ or B2 alloy structures respectively.

\begin{figure}[ht]
\protect{\includegraphics[width=0.8\columnwidth]{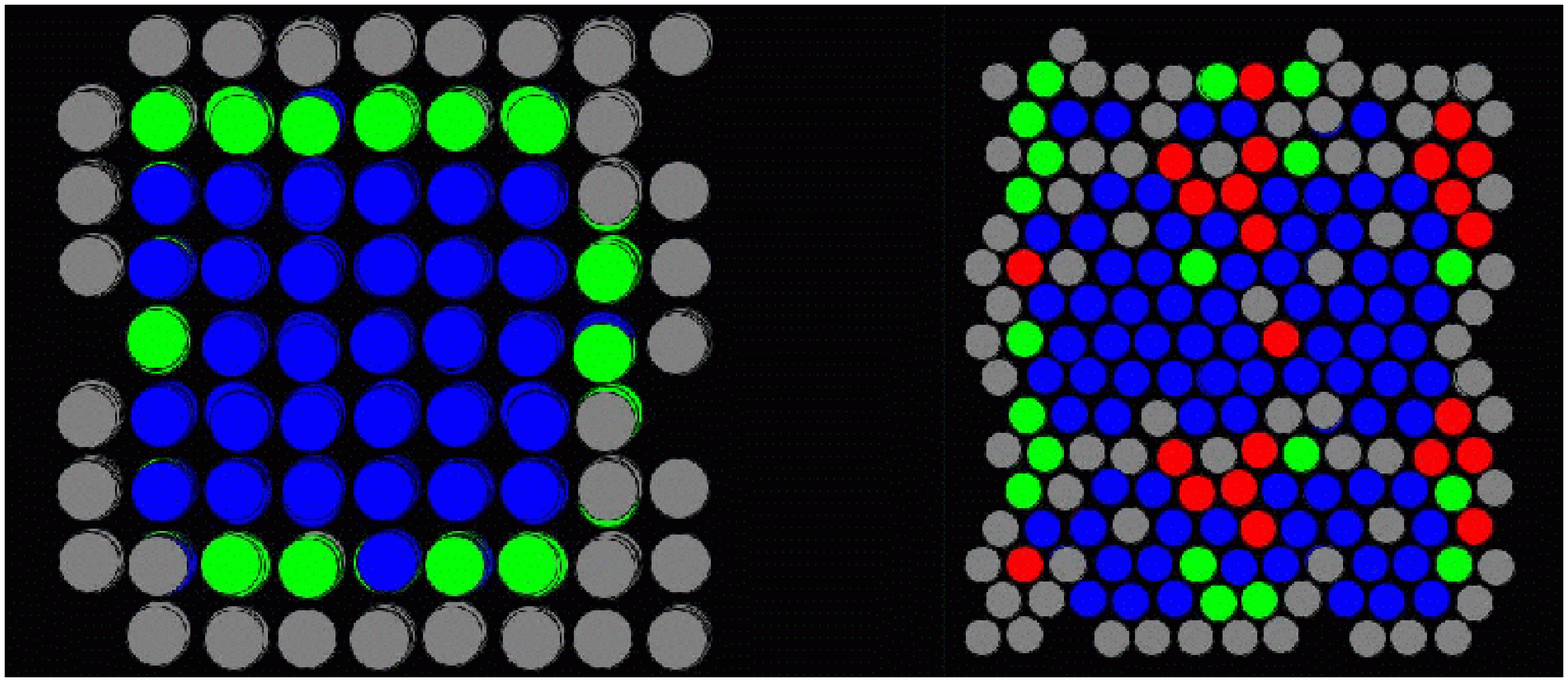}}
\caption{ Local crystal structures in NiTi, derived from cooling ab
initio molecular dynamics from 400K to 200K and quenching to 0K. left: 64 atom simulation viewed along (001) right: 96 atom simulation viewed along (111).
Blue
atoms denote bcc, green fcc, red hcp, grey boundary. In each case the simulation cell is doubled -  periodic boundary
conditions are not applied in the sample boundary, so the outer two
layers of particles are not reliably identified, but every atom in the simulation is correctly identified with all its neighbours in the central region of the figure.
\protect\label{dft}}
\end{figure}

The 64 atom simulation (figure.\ref{dft}) remained almost entirely to
a single phase strained bcc structure.  The 96 atom cell, despite its
high aspect ratio, was also too small to show any significant change
from distorted bcc.  The fully quenched structure has not gone back to B2: 
there is some evidence of a transformation
towards an L1$_{0}$ or B19 structure in a few atoms, but  the best description of the configuration is as distorted B2.

As with the metastable fcc phase in zirconium, the simulations of NiTi
show that nanosized structures can impose stabilizing strains on the
existing structure.  Moreover, the formation energy of its boundary
may be high enough to to alter the stable crystal structure in
microstructures aat a given lengthscale.

\section{Conclusions}

In conclusion, we have shown that in going from the atomistic scale to
the continuum in describing martensitic microstructures the following
features are important:

\begin{itemize}

\item The relative orientation of twins is defined by crystal
symmetry, and will not normally coincide with a low energy twin
boundary.  The twin boundary might be a low-energy boundary with an
array of geometrically necessary dislocations, but in general these
dislocations will be sessile.

\item Some geometrically-allowed boundaries between twins may not appear
because of their high energy.

\item Sessile twin-boundary dislocations may act as stress
concentrators, and thence as sources for mobile dislocations which
allow the boundary to advance.

\item Nanoscale microstructures may stabilise different crystal
structures from the thermodynamic limit, either inhibiting their
appearance through the energy cost of twin boundaries, or allowing an
altogether different phase with lower interface costs to nucleate.

\item Twin boundary energy and mobility varies sharply depending on
the boundary orientation.  Continuum models need to take this into
account.

\end{itemize}

The effects of nanostructure in stabilizing certain crystal structures
outwith their thermodynamic stability ranges offers an resolution to a
number of phenomena observed experimentally but ignored in phase-field
and similar models. The nanostructure creates a variety of local
environments subject to differing inhomogeneous stress whose
minimisation calls for a variety of required interfaces. In particular
metastable ``intermediate'' phases may be stabilised by particularly
favourable local regions of stress and interface energies low
energies.  Moreover, the relative stability of austenite and
martensite varies in different regions of the crystal depending 
of the local stress environment.  Hence both can coexist at the 
same macroscopic (i.e. average) conditions of temperature 
and pressure.  This explains at the microscopic level the existence 
of distinct ``martensite start'' and ``martensite finish'' temperatures.

There remains a need to simulate a binary alloy undergoing a B2-B19'
transition in order to understand the atomic level behaviour of the
shape memory effect.  There remains a discrepancy between the
theoretical perfect crystal structures for NiTi martensite, and
experimental measurement.  This discrepancy is serious, in that the
theoretical structure does not even lie in the EPN of B2.  A
resolution is offered that the theoretical structure is very soft in
shear\cite{xhuang} and so small internal stresses will recover the
experimental value.  This offers the possibility that the shape memory
effect is in part due to microstructural stress, which offers a
mechanism for the two-way effect.

{\bf Acknowledgements:} The author would like to thank colleagues
U.Pinsook, K.Rabe and X.Huang for assistence with the simulations
described here, and the organisers and other participants at PTM2005
for stimulation in writing this paper.  This paper discusses issues
across a wide range of methodologies, and author would like to
apologise for his ignorance of some areas of the literature.


\begin{thebibliography}{99}
\bibitem{glass} The entropy of a glass remains problematic.  In
particular, should one include the configurational entropy of all
possible glass structures as in the liquid.  The argument for doing so
is that on glassification the system could choose any such structure -
a large phase space volume, however once the glassification process is
complete the system is frozen into one small region of phase space,
and cannot sample "equivalent" structures in the way statistical
mechanics assumes.
\bibitem{wrongstack} The zirconium potential used in this work gives far too low a stacking fault energy, so this dissociation may not occur in real zirconium.
\bibitem{DD} There is some confusion as to the precise meaning of
displacive and diffusive at the atomic level.  Here, I take displacive
to mean that each atom in one structure can be mapped to a particular
site in the other.  This does not preclude transformations involving
shuffles, or those proceeding by nucleation and growth.
\end{thebibliography}
\end{document}